\documentclass[12pt]{article}
\usepackage{amsmath, amssymb}
\usepackage{hyperref} 
\usepackage{graphicx}
\usepackage[affil-it]{authblk}
\usepackage[a4paper, total={6.2in, 9.6in}]{geometry}

\newcommand{\Tr}{{\rm Tr}}
\newcommand{\Q}{W} 
\newcommand{\om}{\omega}
\newcommand{\ga}{\gamma}
\newcommand{\E}{E} 
\newcommand{\cL}{{\mathcal L}}

\newtheorem{theorem}{Theorem}

\begin{document}

\title{\LARGE\textbf{Energy control in a quantum oscillator using coherent control and engineered environment}}
	
\author[1,2,*]{Alexander N. Pechen}
\author[3,4,**]{Sergey Borisenok}
\author[5,6,***]{Alexander~L.~Fradkov}

\affil[1]{\normalsize The National University of Science and Technology MISIS,\par
	6~Leninskiy prosp., Moscow 119991, Russia}
\affil[2]{\normalsize Department of Mathematical Methods for Quantum Technologies,\par
	Steklov Mathematical Institute of Russian Academy of Sciences,\par
	8~Gubkina str., Moscow 119991, Russia}
\affil[3]{\normalsize Department of Electrical and Electronics Engineering, Faculty of Engineering, \par
	Abdullah G\"ul University, S\"{u}mer Campus, Kayseri 38080, T\"{u}rkiye}
\affil[4]{\normalsize T\"{u}rkiye Feza G\"{u}rsey Center for Physics and Mathematics, \par
	Bo\u{g}azi{c}i 	University, Kandilli Campus,  Istanbul 34684, T\"{u}rkiye}
\affil[5]{Institute for Problems of Mechanical Engineering of Russian Academy of\par
	 Sciences, V.O., Bolshoj pr. 61, St. Petersburg 199178, Russia}
\affil[6]{Saint Petersburg University,	28 Universitetskii prospect,\par
 Peterhof,  St. Petersburg 195904, Russia}

\affil[*]{apechen@gmail.com, \href{http://www.mathnet.ru/eng/person17991}{mathnet.ru/eng/person17991}}
\affil[**]{sergey.borisenok@agu.edu.tr, \href{http://people.agu.edu.tr/sergeyborisenok}{people.agu.edu.tr/sergeyborisenok}}
\affil[***]{fradkov@mail.com, Corresponding author}

\date{}
\maketitle

\begin{abstract}
We develop and analyze a new method for manipulation of energy in a quantum harmonic oscillator using coherent, e.g., electromagnetic, field and incoherent control. Coherent control is typically implemented by shaped laser pulse or tailored electromagnetic field. Incoherent control is implemented by engineered environment, whose mean number of excitations at the frequency of the oscillator is used as a control variable. An approach to  coherent and incoherent controls design based on the speed gradient algorithms in  general, finite and differential forms is proposed. It is proved that the differential form is able to completely manipulate the energy of the oscillator: an arbitrary energy can be achieved starting from any initial state of the oscillator. The key instrument which allows for complete energy manipulation in this case is the use of the engineered environment. A robustified speed-gradient control algorithm in differential form is also proposed. It is shown that the proposed robustified control algorithm ensures  exponential stability of the closed loop system which is preserved for sampled-data control.
\end{abstract}

\medskip

Highlights:
\begin{itemize}
	\item Coherent and incoherent control of energy of a quantum oscillator is studied.
	\item General, differential and finite forms of the speed gradient algorithm are developed. 
	\item We prove that the differential form can steer energy of the oscillator to any value. 
	\item Feasibility of the energy control in a quantum oscillator is rigorously proven.
	\item A method to explicitly design physical coherent and incoherent controls is provided.
	\item An important problem of sampled-data control is addressed. 
	\item A robustified speed-gradient control algorithm in differential form is proposed. 
	\item This algorithm ensures exponential stability of the closed loop system. 
	\item Simulation results confirm reasonable performance of the closed loop systems.
\end{itemize}

\section{Introduction}

Control of atomic and molecular quantum systems is an emerging branch of modern science~\cite{Koch2022}. An explosion of interest in quantum control happened in the end of the 1980s -- beginning of the 1990s with the rapid development of laser industry which led to creation of ultrafast femtosecond lasers used as a tool to manipulate single atoms and molecules. Quantum control has nowadays a wide range of potential and perspective applications in quantum computing, laser chemistry, quantum technologies, including nanomotors, nanowires,  nanochips, nanorobots, etc.~\cite{RiceBook2000,BrumerBook2003,TannorBook2007,FradkovBook2007,WisemanBook2010,Petersen2010,Moore2011,RabitsReview2012,Glaser2015Report,Lyakhov2020}.

A strong interest in this field is directed towards development of algorithms for efficient manipulation of quantum systems.
Various algorithms were proposed for controlling quantum systems
including gradient ascent pulse engineering (GRAPE)~\cite{GRAPE},
optimal control methods~\cite{Tannor1992,Krotov2009,Morzhin2029,James2021}, the
Broyden-Fletcher-Goldfarb-Shannon (BFGS) algorithm and its
modifications~\cite{BFGS}, Lyapunov-based techniques \cite{Kuang08,Cong2013,Kuang17,Cong2020},
machine learning~\cite{Sanders2011},
speed gradient method~\cite{Fradkov2005,BFP10,PechenBorisenok2015}, chopped random-basis quantum optimization (CRAB)~\cite{CanevaPRA2011}, differentiable programming method~\cite{Schafer2020}, genetic algorithms and evolutionary
strategies~\cite{Judson2002,PeRa06}, and combined
approaches~\cite{Eitan2011,Machnes2011,Goerz2015,Alkhawaja2009,Andreev2019}. Optimal control for cooling of a harmonic oscillator was studied in~\cite{Hoffmann2011}. Typically in the existing works only coherent control is employed \cite{Whaley2015,Vuglar2017,Xiang2017}.

In this work a possibility to achieve efficient energy control in a quantum oscillator  driven both by coherent field and by the engineered environment is analyzed. Compared to the problems with coherent control, an additional control variable --- spectral density at the frequency of the oscillator is introduced. We develop for this control problem the speed-gradient method and apply it to find the shape of the coherent control and desired time dependence of the spectral density of the bath which drive the oscillator to a target energy value. The speed gradient algorithm \cite{FradkovBook2007,AndrFr2021} is used in its general, finite and differential forms. The differential form, as opposite to the finite form, is found to be able to efficiently steer the energy of the oscillator to a predefined target value, both for heating and for cooling the oscillator.

We find that the key factor which allows for the efficient energy transfer in the oscillator is the use of the engineered environment. Engineered environments are used for manipulation of quantum systems in various contexts. They were suggested for improving quantum state engineering~\cite{Cirac2009},
cooling of translational motion without reliance on internal degrees of freedom~\cite{Calarco2011}, etc. Control by electromagnetic fields of quantum systems with infinite dimensional Hilbert space and a discrete spectrum was studied~\cite{Assemat2014}. The possibility to prepare several classes of quantum states of a harmonic oscillator by means of optimal control theory based on Krotov's method was shown~\cite{Rojan2014}. A scheme for steady-state preparation of a harmonic oscillator in the first excited state using a dissipative dynamics which autonomously stabilizes a harmonic oscillator in the $n=1$ Fock state was proposed in~\cite{Borkje2014}. Quantum harmonic oscillator state preparation by reservoir engineering was considered in~\cite{Kienzler2015}. Optimal control for cooling a quantum harmonic oscillator by controlling its frequency was proposed in~\cite{Salamon2012}. Reachable states for a qubit driven by coherent and incoherent controls were analytically described using geometric control theory~\cite{Lokutsievskiy2021}.

This work exploits incoherent control by engineered
environment using spectral density of the bath~\cite{PeRa06,Pe11,MorzhinLJM2021,Petruhanov2022}. The oscillator has equidistant spectrum with only one transition frequency so that in the first order in the system-bath interaction excitations only at this frequency affect the system. At higher orders in the system-bath interaction, an exchange of energy between the system and the bath may involve several bath modes whose excitation energies add up to the system frequency. We neglect it here and exploit only the first order resonant incoherent control considering bath excitations only at the resonant frequency.

Such incoherent control in combination with coherent control was earlier shown to be sufficient to approximately produce arbitrary mixed density matrices of non-degenerate finite level open quantum systems thereby providing their complete controllability~\cite{Pe11}. It is important to note that the method developed in~\cite{Pe11} requires the assumption that the system spectrum is non-degenerate, so that all system's transition frequencies are different. For the case of quantum harmonic oscillator, where transition frequencies between nearest levels are all the same, the method of ~\cite{Pe11} is not applicable i.e. for quantum harmonic oscillator new methods need to be developed.

In this paper we develop a method to control quantum  harmonic oscillator using a combination of coherent and incoherent controls and speed gradient algorithm. It is proved that this method can realize complete manipulation of the energy of the quantum oscillator. The proof is performed using quadratic Lyapunov function. The paper is structured as follows. In Section 2 initial model of the controlled quantum  harmonic oscillator and the control objective are formulated. Section 3 is devoted to formulation of the three-dimensional model for dynamics of average variables as well  as to the exposition of the main results: derivation of the control algorithms and analysis of the closed loop system stability and the control objective achievement by combination of coherent and incoherent controls. System based on incoherent control only is studied in Section 4. In Section 5 the issues of sample data control of quantum systems are discussed.
Simulation results illustrating dynamics of the proposed control systems are presented in Section 6.

\section{Quantum  oscillator in the Markovian bath}
Consider a one-dimensional quantum harmonic oscillator with frequency $\omega_0$. Let $\hat Q$ and $\hat P$ be position and momentum operators of the oscillator which satisfy the canonical commutation relations $[\hat Q,\hat P]=i$ (we set Planck constant $\hbar=1$). Position and momentum operators can be expressed in terms of creation and annihilation operators $\hat a^+$ and $\hat a$ as
\[
 \hat Q=\frac{\hat a+\hat a^+}{\sqrt{2\omega_0}},\qquad \hat P=\sqrt{\frac{\omega_0}{2}}\frac{\hat a-\hat a^+}{i}\, .
\]
The free Hamiltonian of the oscillator is
\[
H_0=\frac{\hat P^2+\omega_0^2\hat Q^2}{2}=\Bigl(\omega_0+\frac{1}{2}\Bigr) \hat a^+\hat a \, .
\]

If the oscillator is coupled to an environment, then its state at time $t$ is described by density matrix $\rho_t$, which is a positive operator with unit trace, $\Tr\rho_t=1$. If the oscillator is coupled to a Markovian bath, then dynamics of its density matrix are governed by  Markovian master
equation of Kossakowski-Gorini-Sudarschan-Lindblad form
\begin{equation}\label{eq1}
\frac{d\rho_t}{dt}=-i[H_0+u(t)\hat Q,\rho_t]+\cL(\rho_t)\, .
\end{equation}
Here real function $u(t)$ is coherent control and the dissipative generator has the following form
\begin{eqnarray}
\cL(\rho_t)=\gamma (n(t)+1)(2\hat a\rho \hat a^+-\rho \hat a^+\hat a-\hat a^+\hat a\rho)+\gamma n(t)(2\hat a^+\rho \hat a-\hat a\hat a^+\rho-\rho \hat a\hat a^+)\, , \label{eq1.2}
\end{eqnarray}
where $\gamma>0$ is the relaxation rate and  $n(t)\ge 0.$ Master equation (\ref{eq1}) with dissipative generator (\ref{eq1.2})for time independent Hamiltonian $H$ and spectral density $n$ was considered in \cite{BreuerBook}. Inspired by  \cite{PeRa06}, we include coherent control $u(t)$ in the Hamiltonian $H$ and consider  time dependent spectral density $n(t)$ as an incoherent control.
 Physically $n(t)$ is the mean number of excitations in the bath at energy $\om_0$ and at time $t$. If the bath is at the inverse temperature $\beta(t)=1/T(t)$, where $T(t)$ is the temperature of the bath at time $t$, then its spectral density at $\omega_0$ is
\[
 n(t)=\frac{1}{e^{\om_0\beta(t)}-1} \, .
\]
In general, bath can be non-thermal with arbitrary spectral density. Incoherent control by spectral density of the bath was used for manipulation by density matrices in finite level atomic systems~\cite{PeRa06} and was shown, in combination with coherent control, to be rich enough to realize complete density matrix controllability for a wide class of finite level quantum systems with non-degenerate spectrum (recall that quantum harmonic oscillator is beyond that class)~\cite{Pe11}.

The most natural example of the bath for this method is the bath of incoherent photons. The average excitation in the bath in this case is just the number of photons with frequency $\omega_0$. Then to realize the obtained scheme, we do not need to be able to modify or establish a thermal equilibrium of the full state of the bath. We just need a sufficiently fast source for incoherent pumping of photons at frequency $\omega_0$.

Average energy of the oscillator is $\E(\rho)=\Tr(H_0\rho)$. Our goal is to develop a method to find time dependent controls $u(t)$ and $n(t)$ (or, equivalently, $u(t)$ and $\beta(t)$), such that energy of the oscillator driven by these controls converges to the predefined value $E_*$.

The problem of heating or cooling the oscillator towards a prespecified value of energy $E_*$ can be formulated as minimization of the objective function
\begin{equation}
\label{goal}
 W_t\to 0 \textrm{ as } t\to\infty\, .
\end{equation}
where $W_t=W(E(t)), ~~W(E)=\frac{1}{2}(E-E_*)^2$. Apparently the objective function $W(E)$ is non-negative, $W(E)\ge 0$ and satisfies $W(E)=0$ if and only if $E=E_*$.

\section{Heat transfer by combination of coherent and incoherent controls}

In this section we consider the case when both coherent and incoherent controls are used to manipulate the energy in quantum  oscillator. Let $Q_t=\Tr(\hat Q\rho_t)$ and $P_t=\Tr(\hat P\rho_t)$ be average values of position and momentum at time $t$. Direct computations yield
\begin{eqnarray*}
 \Tr (\hat a\cL(\rho_t))=-\ga\Tr (\hat a\rho_t),\qquad \Tr (\hat a^+\cL(\rho_t))=-\ga\Tr (\hat a^+\rho_t)\, .
\end{eqnarray*}
Therefore
\begin{eqnarray*}
 \Tr (\hat Q\cL(\rho_t))&=&-\ga Q\, ,\\
 \Tr (\hat P\cL(\rho_t))&=&-\ga P\, .
\end{eqnarray*}
This gives the following  system of equations
\begin{eqnarray}
 \frac{dE}{dt}&=&-u(t)P+2\ga(\om_0n(t)-E)\label{e1}\\
 \frac{dQ}{dt}&=&P-\ga Q\label{e2}\\
 \frac{dP}{dt}&=&-\omega_0^2 Q-u(t)-\ga P\, .\label{e3}
\end{eqnarray}
Thus the model of the quantum oscillator is reduced to the system of equations (\ref{e1})-(\ref{e3}) describing dynamics of the average energy, position and momentum of the quantum  oscillator under the action of coherent and incoherent controls $u(t), n(t)$.

\subsection{The speed gradient control algorithms}
To design the control algorithm we use the speed gradient method~\cite{FradkovBook2007,AndrFr2021}. Speed-gradient method is closely related to Lyapunov design methods \cite{Kuang08,Cong2013,Kuang17}. The difference is in the methodology. According to the speed-gradient methodology the control goal should be represented as asymptotic minimization of the goal (objective) function. Then one needs to evaluate the speed of changing the objective function $W$ along the trajectories of the controlled system and then evaluate the gradient of this speed  with respect to the control vector. In our case the speed of changing the objective function $W_t$ along solution of (\ref{e1})-(\ref{e3}) has the form
\begin{eqnarray}
\om(u,n)=\frac{\partial  \Q_t}{\partial{\E}} \frac{d\E}{dt}=(\E-E_*)(-uP+2\gamma (\omega_0 n-\E))\, .\label{speed}
\end{eqnarray}
Its gradient with respect to controls is
\begin{eqnarray}
\nabla_{u}\om(u,n)&=&-P(\E -E_*)\\
\nabla_{n}\om(u,n)&=&2\gamma\omega_0(\E -E_*)\, .
\end{eqnarray}
This leads to the following general (finite-differential) form of the speed gradient algorithm for transfer of heat in quantum oscillator by coherent and incoherent controls
\begin{eqnarray}\label{2.5}
 \frac{d}{dt}[u+\psi_1(u,n)]&=&-\Gamma_1\nabla_u\om(u,n)=\Gamma_1 P(\E -E_*)\\
 \frac{d}{dt}[n+\psi_2(u,n)]&=&-\Gamma_2\nabla_n\om(u,n)=-\tilde\Gamma_2 (\E -E_*)\, .
\end{eqnarray}
Here $\Gamma_{1,2}\ge 0$, $\tilde\Gamma_2=2\gamma\omega_0 \Gamma_2\ge 0$, and vectors $\psi_{1,2}$ satisfy the pseudogradient conditions $\langle\psi_1,\nabla_u\om\rangle\ge 0$ and $\langle\psi_2,\nabla_n\om\rangle\ge 0$.

Special cases of (\ref{2.5}) are the differential form and the finite form of the algorithm.

{\bf SGA-D:} {\it
Differential form of the speed gradient algorithm for transfer of heat in quantum harmonic oscillator using both coherent and incoherent control is
\begin{eqnarray}
 \frac{du}{dt}&=&\Gamma_1 P(\E -E_*)\label{2.6}\\
 \frac{dn}{dt}&=&-\tilde\Gamma_2 (\E -E_*)\, .\label{2.6.2}
\end{eqnarray}
}

{\bf SGA-F:} {\it
Finite (linear) form of the speed gradient algorithm for transfer of heat in quantum harmonic oscillator using only incoherent control is
\begin{eqnarray}\label{2.7}
 u&=&\Gamma_1 P(\E -E_*)\\
 n&=&-\tilde\Gamma_2 (\E -E_*)\, . \label{2.7.2}
\end{eqnarray}
}

\subsection{Convergence of the differential form of the control algorithm}
Although the proposed algorithms are designed according to the general speed-gradient design procedure, the convergence of the algorithm (achievement of the control objective and boundedness of all the closed loop system variables) do not follow from the existing general statements about convergence of the speed-gradient algorithms \cite{FradkovBook2007,AndrFr2021}. The reason is in that the objective function $W(E)=\frac{1}{2}(E-E_*)^2$ is not proper for system (\ref{e1})-(\ref{e3}). Besides it does not possess radial unboundedness which is standard property to prove boundedness of the closed loop trajectories. Therefore we need to examine the closed loop system dynamical  properties which will be done now.

First we will prove that the differential form of the speed gradient algorithm  ensures convergence of the system energy $E(t)$ to the target energy value $E_*$ for any initial conditions.

\begin{theorem}
Consider quantum harmonic oscillator (\ref{e1})-(\ref{e3}) evolving under the action of coherent control $u(t)$ and incoherent control $n(t)$
satisfying ~(\ref{2.6}) and~(\ref{2.6.2}). Then all the trajectories of the closed loop system are bounded and the average energy of the oscillator asymptotically converges to $E_*$, $\lim\limits_{t\to+\infty} E(t)=E_*$.
\end{theorem}
{\bf Proof.} Consider Lyapunov function in the form
\begin{equation}\label{eq:Lyapunov}
V_1(E,u,n)=W(E)+\frac{1}{2\Gamma_1}(u-u_*)^2+\frac{1}{2\Gamma_2}(n-n_*)^2\, .
\end{equation}
Here $u_*$ and $n_*$ are some constants that  will be chosen later to make $\dot V_1\le 0$.
Using equations~(\ref{e1})--(\ref{e3}),~(\ref{2.6}) and~(\ref{2.6.2}) yields
\begin{eqnarray*}
\dot V_1&=&\vphantom{\frac{\tilde \Gamma_2}{\Gamma_2}}
\dot W+\frac{1}{\Gamma_1}(u-u_*)\dot u+\frac{1}{\Gamma_2}(n-n_*)\dot n\\
&=& \vphantom{\frac{\tilde \Gamma_2}{\Gamma_2}}
(E-E_*)\dot E+(u-u_*)P(E-E_*)-\frac{\tilde \Gamma_2}{\Gamma_2}(n-n_*)(E-E_*)\\
&=& \vphantom{\frac{\tilde \Gamma_2}{\Gamma_2}}
(E-E_*)\bigl(-uP+2\gamma(\omega_0 n-E)+(u-u_*)P-\frac{\tilde \Gamma_2}{\Gamma_2}(n-n_*)\bigr)\\
&=& \vphantom{\frac{\tilde \Gamma_2}{\Gamma_2}}
(E-E_*)(-2\gamma E-u_*P+2\gamma\omega_0 n_*)\, .
\end{eqnarray*}
Recall that  $\Gamma_2=\tilde\Gamma_2/2\gamma\omega_0$ and set
\[
u_*=0\, , \qquad n_*=\frac{E_*}{\omega_0}\, .
\]
Then
\[
\dot V_1=-2\gamma(E-E_*)^2=-\gamma W\le 0\, .
\]
Denote $V_{1t}=V_1(E(t),u(t),n(t))$. Then $V_{1t}\le V_0<\infty$ and therefore functions $E(t)$, $u(t)$, and $n(t)$ are bounded on the interval $[0,t*)$ where solution of equations~(\ref{e1})--(\ref{e3}),~(\ref{2.6}) and~(\ref{2.6.2}) exists. Indeed, according to~(\ref{eq:Lyapunov}) the Lyapunov function is the sum of three non-negative terms. Boundedness of $V_{1}(t)$ implies boundedness of all three functions $E(t),u(t),n(t)$.

However boundedness of $V_{1}(t)$ does not imply immediately boundedness of  $P(t)$ and $Q(t)$. To prove that state vector is bounded and the system is  forward complete ($t*=\infty$), denote
\[
X=\left(\begin{array}{c}
P\\
Q
\end{array}\right),\qquad
A=\left(\begin{array}{cc}
-\gamma & -\omega_0^2\\
1 & -\gamma
\end{array}\right),\qquad
B=\left(\begin{array}{c}
-1\\
0
\end{array}\right)\,.
\]
Then (\ref{e2}),(\ref{e3}) may be rewritten as
\begin{equation}\label{PQ}
\dot X=AX+Bu\, ,
\end{equation}
where matrix $A$ has eigenvalues $\lambda_{1,2}=-\gamma\pm\omega_0$ such that ${\rm Re}\lambda_{1,2}=-\gamma<0$, i.e. matrix $A$ is Hurwitz. Since $u(t)$ is bounded,  for initial conditions $P(0),Q(0)$ from a bounded set functions $P(t),Q(t)$ are bounded for $0\le t\le t*$.  Hence $X(t)$ is bounded for all $t\ge0$ t.e. $t*=\infty$  and the system is forward complete. Boundedness of all variables implies that $\dot V_1(t)$ is uniformly continuous. Hence by Barbalat's lemma $\dot V\to 0$ as $t\to\infty$ and therefore $W(E(t))\to 0$. This proves that the speed gradient algorithm ensures convergence of the system energy function to the target energy value and finishes the proof of the theorem.

\subsection{Conditions for non-negativity of the environmental spectral density $n(t)$}
A priori the speed gradient method does not guarantee non-negativity of the designed control function $n(t)$. However, by its physical meaning spectral density $n(t)$ must be non-negative and, as will be shown below, it puts some restrictions on the algorithm parameter $\Gamma_2$.  The cases of heating $E_*>E(0)$ and cooling $E_*<E(0)$ will be examined separately. It is convenient to introduce dimensionless quantity $\alpha=\sqrt{\Gamma_2}\omega_0$.

\begin{theorem} 1. In the case of heating $E_*>E(0)$ for non-negativity of $n(t)$ it is sufficient to choose $\alpha\le 1$, or, equivalently,
\begin{equation}\label{Positivity}
\Gamma_2\le \frac{1}{\omega_0^2}\, .
\end{equation}
2. In the case of cooling $E_*<E(0)$ for positivity of $n(t)$ it is sufficient to choose
\[
\alpha\le \frac{1}{\frac{E(0)}{E_*}-1}\, ,
\]
or, equivalently,
\begin{equation}
\Gamma_2\le \frac{1}{\omega_0^2\left(\frac{E(0)}{E_*}-1\right)^2}\, .
\end{equation}
\end{theorem}
{\bf Proof.} Recall that since $u_*=0$, Lyapunov function~(\ref{eq:Lyapunov}) has the form
\[
 V(E,u,n)=W(E)+\frac{1}{2\Gamma_1}u^2+\frac{1}{2\Gamma_2}(n-n_*)^2\, .
\]
Using the relation $V_t\le V_0$, we obtain
\begin{eqnarray*}
 W((E(t))+\frac{1}{2\Gamma_1}u(t)^2+\frac{1}{2\Gamma_2}(n(t)-n_*)^2
 \le V_0\, .
\end{eqnarray*}
Therefore
\begin{eqnarray*}
 (n(t)-n_*)^2\le 2\Gamma_2 V_0\, .
\end{eqnarray*}
Hence
\[
 n_*-\sqrt{2\Gamma_2 V_0}\le n(t)\le n_*+\sqrt{2\Gamma_2 V_0}\, .
\]
To guarantee nonnegativity of $n(t)$, it is sufficient to choose
\begin{equation}\label{eq2}
 n_*-\sqrt{2\Gamma_2 V_0}\ge 0\, .
\end{equation}
Let us choose the initial conditions $u(0)$ and $n(0)$ in order to minimize $V_0$. This requires to set
\begin{equation*}
 u(0)=0,\qquad n(0)=n_*= \frac{E_*}{\omega_0}\, .
\end{equation*}
Then
\[
 V_0=W(E(0))=\frac{1}{2} (E(0)-E_*)^2
\]
and sufficient condition~(\ref{eq2}) for $n(t)\ge 0$ becomes
\begin{equation*}
 E_*\ge \alpha|E(0)-E_*|\,.
\end{equation*}

First consider the case of heating, so that $E_*>E(0)$. If $\alpha\le 1$ then heating to any energy $E_*>E(0)$ is possible. Hence for heating to any energy it is sufficient to choose
\begin{equation}\label{Positivity2}
\Gamma_2\le \frac{1}{\omega_0^2}\, .
\end{equation}

Next let us consider the case of cooling, so that $E_*<E(0)$. In this case
\[
 E_*\ge \frac{\alpha}{1+\alpha}E(0)=\delta(\alpha) E(0), \quad \textrm{where}\quad \delta(\alpha)=\frac{\alpha}{1+\alpha}\, .
\]
It means that cooling is possible down to the energy level $\delta(\alpha) E(0)$. Given value of $E_*$, the value of $\alpha$ has to satisfy $\delta(\alpha)\le  E_*/E(0)$ which in turn gives the second statement of the theorem and finishes its proof.

The conditions of the Theorem suggest that for cooling to a low energy the value of $\alpha$, or equivalently, $\Gamma_2$ should be sufficiently small. We remark however that these conditions are sufficient but may be not necessary.

\section{Heat transfer by incoherent control only}
In this section we develop the control algorithm for heat transfer in the oscillator when coherent control is switched off ($u(t)=0$), and only incoherent control $n(t)\ge 0$ is used to manipulate with the density matrix of the oscillator.

For $u(t)=0$ equation (\ref{e1}) containing control $n(t)$ becomes separated from (\ref{e2})-(\ref{e3}) and reads
\begin{equation}\label{eqE}
 \frac{d\E}{dt}=2\gamma(\omega_0 n-\E)\, .
\end{equation}
Since the subsystem (\ref{e2})-(\ref{e3}) is asymptotically stable, it is sufficient to stabilize the equation (\ref{eqE}).
With the aim of reducing complexity of the control system let us first design the speed-gradient control in the finite form (\ref{2.7.2})
Evaluation of the speed of changing the objective function $W(E)=\frac{1}{2}(E-E_*)^2$ along the system trajectory yields
\begin{equation}
\om(n)=\frac{\partial  W(E)}{\partial{\E}} \frac{d\E}{dt}=2\gamma (\E-E_*)(\omega_0 n-\E)\, .
\end{equation}
The gradient of the speed with respect to control is as follows
\begin{eqnarray}
\nabla_{n}\om(n)=2\gamma\omega_0(\E -E_*)\, .
\end{eqnarray}
Therefore  the finite form of the speed gradient algorithm for transfer of energy in quantum  oscillator by incoherent control is as follows
\begin{equation}\label{eqFinite}
n=-\Gamma(\E-E_*)\, ,
\end{equation}
where $\Gamma>0.$
For finite form the condition $n\ge 0$ implies $\E\le E_*$, i.e. algorithm (\ref{eqFinite}) may work only for heating the system. Substitution of~(\ref{eqFinite}) into~(\ref{eqE}) yields
\begin{equation}\label{eqE2}
 \frac{d\E}{dt}=-2\gamma[\Gamma(\E-E_*)\omega_0+\E]\, .
\end{equation}
The solution of this equation is
\[
 \E(t)=\frac{E_*}{1+\frac{1}{\omega_0\Gamma}}(1-e^{-\Omega t})+E(0)e^{-\Omega t}\, .
\]
Here $\Omega=2\gamma(\omega_0\Gamma+1)$. The limit of this solution as $t\to\infty$ is
\[
  \lim\limits_{t\to+\infty}\E(t)=\frac{E_*}{1+\frac{1}{\omega_0\Gamma}}\, .
\]
The deviation from the target value $E_*$ behaves as $E_*/(\omega_0\Gamma)$. Hence to approach the target energy value $E_*$ as close as possible with the finite form of the speed gradient algorithm and incoherent control one needs to make $\Gamma$ as large as possible.

Let us modify control algorithm to improve asymptotic behavior of the control system. To this end the control value may be found from the equation
$\omega_0 n-\E=-\kappa(E-E_*)$ as follows
\begin{equation}\label{eq8}
n(t)=\frac{(1-\kappa)E+\kappa E_*}{\omega_0}\, .
\end{equation}
Apparently for $0<\kappa<1$ the condition $n(t)\ge 0$ holds for all $E\ge 0, E_*\ge 0$, i.e. algorithm (\ref{eq8}) applies both for heating and for cooling.
Substitution of (\ref{eq8}) into (\ref{eqE}) yields
\begin{equation}\label{eqEa}
 \frac{d\E}{dt}=-2\gamma\kappa(E-E_*)\, ,
\end{equation}
i.e. the goal is achieved with the exponential rate.

\section{Sampled-data control}
The proposed speed gradient algorithms are described in continuous time i.e. they require continuous measurement of some system observables. Non-selective continuous quantum measurements can be also used for quantum control via quantum anti-Zeno dynamics~\cite{AntiZeno1,AntiZeno2}. However, continuous measurements can be difficult to implement in practice, where a finite and relatively small number of measurements can be typically performed. Some results on control with discrete quantum measurements are available in the literature. E.g., general method for controlling quantum systems by back-action of a finite number of discrete quantum measurements was proposed in~\cite{PechenRabitz2006}.
This incoherent control strategy using back-action of quantum measurements was applied to a number of problems: enhancing of the controllability of systems with dynamical symmetry~\cite{Shuang2008}, developing of the incoherent control schemes for quantum systems with wavefunction-controllable subspaces~\cite{DongIEEE2008},
producing sliding mode control for two-level quantum systems~\cite{Dong2012}, manipulating a qubit~\cite{Blok2014}, incoherent control of the retinal isomerization in rhodopsin~\cite{LucasPRL2014}, controlling transitions in the Landau--Zener system~\cite{PechenTrushechkin2015}, inducing quantum Zeno effect in a superconducting flux qubit~\cite{KakuyanagiNJOP2015}, etc.

Let us  examine the sampled-data version of the speed-gradient algorithm {\bf SGA-D}. Sampled-data nonlinear control is studied in a number of publications, see, e.g. \cite{DF81,NesicTeel01,LailaNesicAstolfi06,Hetel17}. In most cases stability and performance of the sampled-data systems are derived from the robustness of their continuous-time versions. It is known (\cite{AndrFr2021} that the speed-gradient algorithm {\bf SGA-D} does not provide appropriate robustness of the closed loop system since it is only marginally stable ($\dot V_1(t)\le 0$) Let us modify control algorithm  to enhance its robustness as follows.

Algorithm {\bf SGA-DR:}
\begin{eqnarray}
 \frac{du}{dt}&=&\Gamma_1 P(\E -E_*)-\alpha_1 u\label{SGA-DR1}\\
 \frac{dn}{dt}&=&-\Gamma_2 (\E -E_*)-\alpha_2(n-n_*)\, .\label{SGA-DR2}
\end{eqnarray}

The key property of the robustified algorithm {\bf SGA-DR} is that it provides exponential stability of the closed loop system.

\begin{theorem} Let $\gamma>4\Gamma_1\Gamma_2$. Then the equilibrium $z*={\rm col}(E*,0,0,0,n_*)$ of the closed loop system (\ref{e1})-(\ref{e3}),(\ref{SGA-DR1}),~(\ref{SGA-DR2}) is exponentially stable.
\end{theorem}

{\bf Proof.} According to Krasovsky theorem, it is sufficient to find a function $V(z)$, where $z={\rm col}(E,P,Q,u,n)$ such that $0\le\kappa_0||z-z*||^2\le V(z)\le\kappa_1||z-z*||^2$ and $\dot V(z)\le -\kappa_2||z-z*||^2$ for some positive $\kappa_0,\kappa_1,\kappa_2.$ Let $A$ be the matrix from (\ref{PQ}) and $R=R^{T}>0$ be the solution of the Lyapunov inequality
\begin{equation}\label{Lyap}
RA+A^{T}R\le -\gamma_0R
\end{equation}
(such a solution exists for $0<\gamma_0<\gamma.$) Let $V_2(X)=X^{T}RX$, where $X={\rm col}(P,Q)$
and $V(z)=V_1(E,u,n)+V_2(P,Q),$ where $V_1$ is from (\ref{eq:Lyapunov}), 
Evaluate the velocity of changing $V(z)$ along trajectories of
(\ref{e1})-(\ref{e3}),(\ref{SGA-DR1}),~(\ref{SGA-DR2}):
\begin{eqnarray*}
\dot V&=&(E-E*)(-2\gamma E+2\gamma E*)+2X^{T}RAX+2X^{T}RBu+\Gamma_1^{-1}u\dot u+\Gamma_2^{-1}(n-n*)\dot n\\
&\le& -2\gamma(E-E*)^2-\gamma_0V_2(X)+2\mu V_2(X)+
2\mu^{-1}u^2-\alpha_1 u^2-\alpha_2(n-n*)^2,
\end{eqnarray*}
where $\mu>0$ is an auxiliary parameter to choose. Choosing $\mu$
satisfying inequalities $\Gamma_2^{-1}\gamma_0-2\mu>0$ and $\Gamma_1^{-1}-2/\mu>0,$
(this is feasible under conditions of the theorem) we obtain $\dot V(z)\le-\alpha V(z),$ where
$$
\alpha={\rm min}\{4\gamma, \Gamma_1^{-1}-2/\mu, \Gamma_2^{-1} \gamma_0-2\mu\}.
$$
Thus the Lyapunov function $V(z)$ fits all conditions of Krasovsky theorem. The theorem is proven.

Now we are in position to formulate the main result of this section concerning convergence of the discretized control algorithm
\begin{eqnarray}
\bar u(t)&=&\bar u(t_k),\bar n(t)=\bar n(t_k), t_k\le t<t_{k+1},\label{SGA-DRS1}\\
 \bar u(t_{k+1})&=&\bar u(t_k)+\Gamma_1 P(t_k)(\E(t_k) -E_*)-\alpha_1 u(t_k)\label{SGA-DRS2}\\
 \bar n(t_{k+1})&=&\bar n(t_k)-\Gamma_2 (\E(t_k) -E_*)-\alpha_2(n(t_k)-n*)\, \label{SGA-DRS3}
\end{eqnarray}
where $t_k=kh.$

{\bf Corollary}. There exists $\bar h>0$ such that for all $0<h<\bar h$ the equilibrium $(E*,0,0,0,n*)$ of the sampled-data closed loop system (\ref{e1})-(\ref{e3}),(\ref{SGA-DRS1})-(\ref{SGA-DRS3}) is exponentially stable.

The statement of the Corollary follows from the Dragan-Khalanai theorem
\cite{DraganKhalanai90} which we formulate here for completeness. Consider a controlled system $\dot x=f(x,u), y=s(x)$, where $x\in R^n, u\in R^m, y\in R^l,$  with a dynamical controller $\dot w=g(w,y), u=U(w,y),$ where $w\in R^p.$ The closed loop system is described by differential equations
\begin{eqnarray}\label{cont-sm}
\dot x=f(x,U(w,s(x))), \dot w=g(w,s(x)).
\end{eqnarray}
Consider a sampled-data controlled system
\begin{eqnarray}
\dot x&=&f(x(t),u(t)), u(t)=u_k, w(t)=w_k, t_k\le t<t_{k+1}\label{disc-sm1}\\
u_k&=&U(w_k,s(x(t_k))), w_{k+1}=w_k+hg(w_k,s(x(t_k))),\label{disc-sm2}
\end{eqnarray}
where $t_k=kh, ~h>0, j=0,1,2,...$

\begin{theorem}[Dragan-Khalanai, 1990]\cite{DraganKhalanai90}. Let functions $f,g,s,U$ be locally Lipschitz and $(\bar x,\bar w)$ be an equilibrium of system (\ref{cont-sm}), i.e. $f(\bar x,U(\bar w,s(\bar x)))=0, g(\bar w,s(\bar x))=0.$ Let there exist
a domain $\mathcal{D}$ containing the point $(\bar x,\bar w)$ and constants $\alpha>0, \beta\ge 1$ such that if $(x(0),w(0))\in \mathcal{D}$, then
solution  $(x(t),w(t))$ of  (\ref{cont-sm}) satisfies for $t\ge 0$ the inequality
\begin{eqnarray}\label{cont-expstab}
||x(t)-\bar x||+||w(t)-\bar w||\le \beta e^{-\alpha t}(||x(0)-\bar x||+||w(0)-\bar w||).
\end{eqnarray}
Let $\mathcal{B}_r=\{(x,w):||x-\bar x||+||w-\bar w||\le r\}$ and $r>0$ satisfies
$\mathcal{B}_r \subset\mathcal{D}.$ Let $L>0$ be an upper bound of Lipschitz constants for all functions $f,g,s,U$ in the compactum $\mathcal{B}_r.$

Then there exist $h>0$ depending on $\alpha, \beta, L$ and  $\tilde\alpha>0, \tilde\beta\ge 1$ such that if $(\tilde x(t),\tilde w(t)), t\ge 0$ is a solution of system (\ref{disc-sm1}),(\ref{disc-sm2}) and $(\tilde x(0),\tilde w(0))
\in \mathcal{B}_{r/2\beta},$ then
\begin{eqnarray}\label{disc-expstab}
||\tilde x(t)-\bar x||+||\tilde w(t)-\bar w||\le \tilde\beta e^{-\tilde\alpha t}(||\tilde x(0)-\bar x||+||\tilde w(0)-\bar w||).
\end{eqnarray}
\end{theorem}
\section{Numerical analysis of the speed gradient quantum control algorithm}
In this section we numerically analyse the speed gradient algorithm in differential and finite forms for heating and cooling of the quantum oscillator affected by both coherent and incoherent controls. We chose the system of units such that $\omega_0=1$. The decoherence rate is $\gamma=1$.

\begin{figure*}
\begin{center}
\includegraphics[width=1\linewidth]{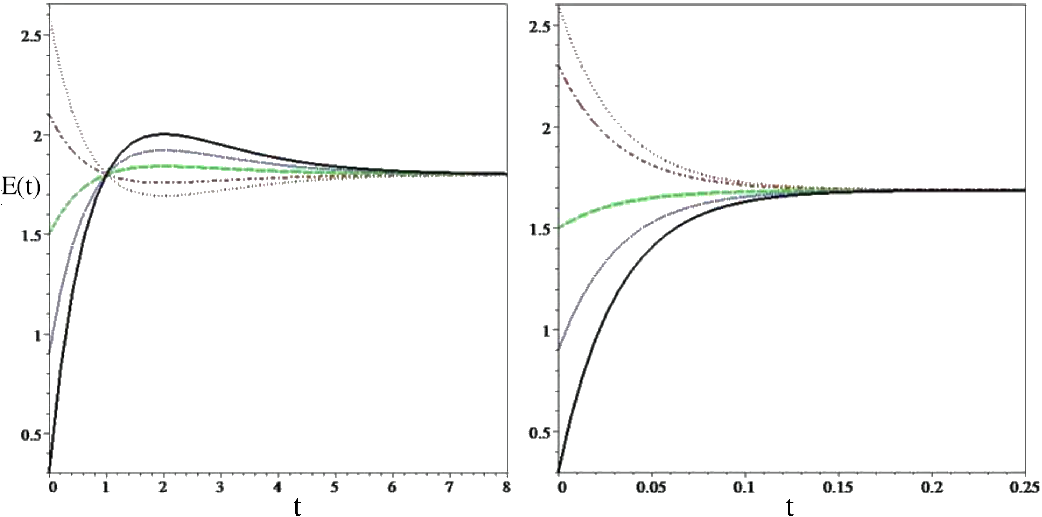}
\caption{Plots of $E(t)$ for heating and cooling of quantum oscillator for different initial values of $E(0)$. Left: control by the differential form of the  algorithm. The parameter values are $\omega_0=1$, $\gamma = 1$, $\Gamma_1 = 3$, $\Gamma_2 = 0.5$, $E_* = 1.8$. Right: Control by the finite form of the  algorithm. The parameter values are $\omega_0=1$, $\gamma = 1$; $\Gamma_1 = 3$, $\Gamma_2 = 15$, $E_* = 0.6$. The plots show good convergence to the target energy value both for heating and for cooling.\label{fig1}}
\end{center}
\end{figure*}

Figure~\ref{fig1} shows the behavior of the energy $E(t)$ for different initial states for heating or cooling the oscillator under the action of coherent and incoherent controls which were constructed  using differential (left) and finite (right) forms of the speed gradient algorithm. The plots show a good convergence to the target energies for all initial values.

\begin{figure*}
\begin{center}
\includegraphics[width=1\linewidth]{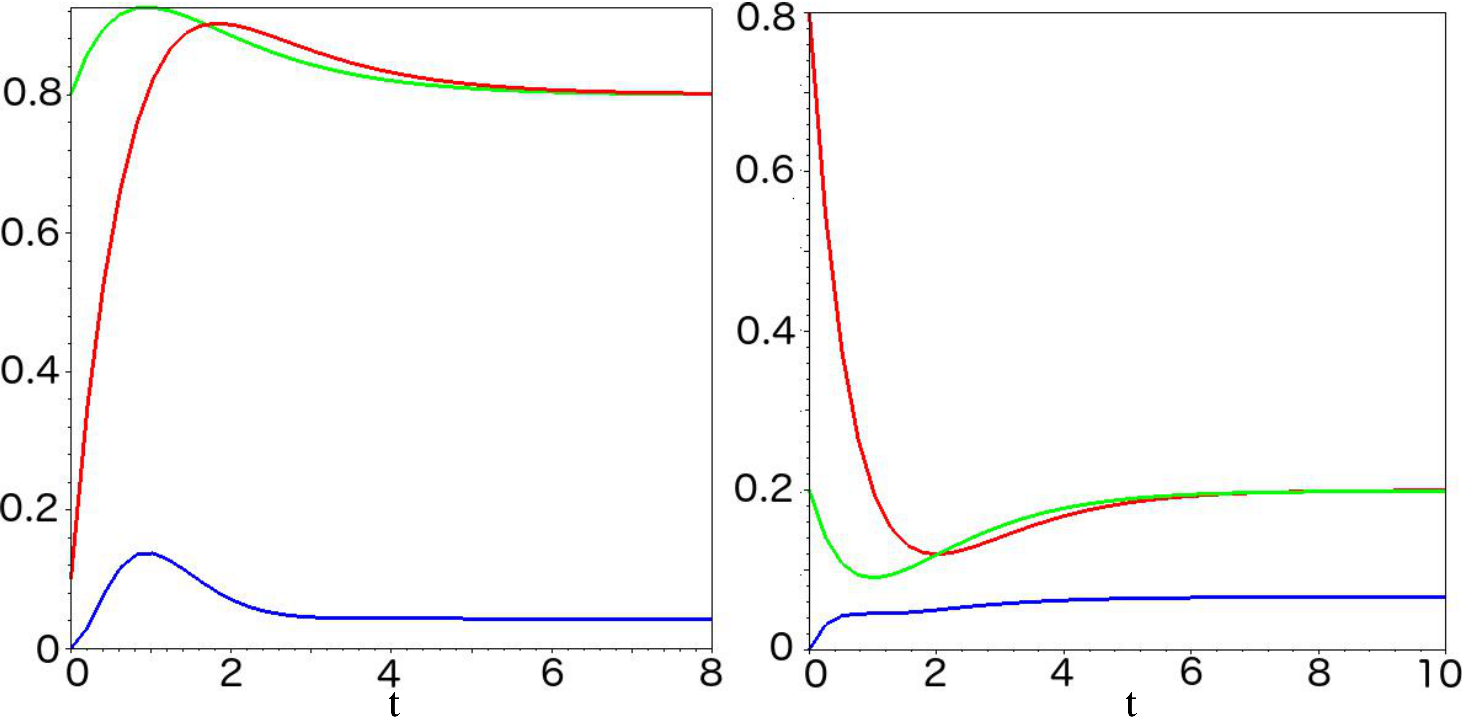}
\caption{Manipulation of the quantum oscillator energy using differential form of the speed gradient algorithm. Left: heating from the initial energy value $E(0)=0.1$ to the target energy value $E_*=0.8$. Right: cooling from the initial energy value $E(0)=0.8$ to the target energy value $E_*=0.2$. The parameters are $\omega_0=1$, $\gamma = 1$; $\Gamma_1 = 3$, $\Gamma_2 = 0.5$. Red lines are for $E(t)$, green for $n(t)$, and blue for $u(t)$.\label{fig2}}
\end{center}
\end{figure*}

Figure~\ref{fig2} shows the behavior of the energy $E(t)$ (red lines) under the action of coherent control $u(t)$ (blue lines) and incoherent control $n(t)$ (green lines) found by the differential form of the speed gradient algorithm~(\ref{2.6}),~(\ref{2.6.2}). Left subplot shows heating from the initial energy value $E(0)=0.2$ to the target energy value $E_*=0.8$. Right subplot shows cooling from the initial energy value $E(0)=0.8$ to the target energy value $E_*=0.3$. The initial conditions are $P(0)=0$ and $ Q(0)=1$.

\begin{figure}
\begin{center}
\includegraphics[width=.7\linewidth]{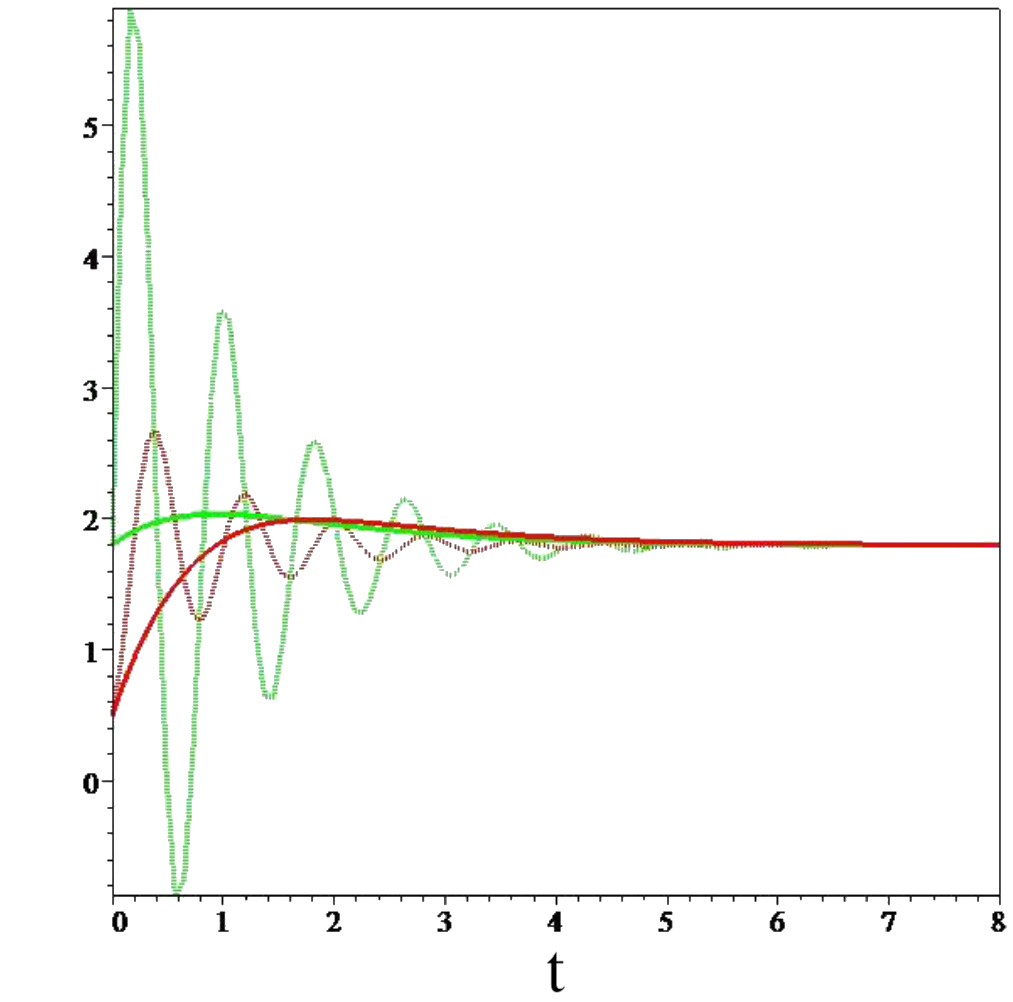}
\caption{Manipulation of the  quantum oscillator energy using differential form of the speed gradient algorithm. Red lines are for $E(t)$, green for $n(t)$. Initial energy value $E(0) = 0.5$ to the target energy value $E_* = 1.8$; $\Gamma _2=0.5$ (the solid lines) and $\Gamma_2=30$ (the dotted lines). The other parameters are $\omega_0 = 1$,  $\gamma= 1$; $\Gamma_ 1 = 3$. This figure illustrates the importance of the condition~(\ref{Positivity}) to guarantee non-negativity of $n(t)$.
\label{fig3}}
\end{center}
\end{figure}

Figure~\ref{fig3} illustrates the importance of the condition~(\ref{Positivity}) to guarantee non-negativity of $n(t)$. The initial energy $E(0)=0.1$ is the same as on the left subplot of Fig.~\ref{fig2}. However, here $\Gamma_2=10>1/\omega_0^2$ and spectral density $n(t)$ (green line) found by the differential form of the algorithm takes at some time instants negative values, in sharp contrast with behavior on Fig.~\ref{fig2} where $n(t)$ (green line) is always positive.

\begin{figure}
\begin{center}
\includegraphics[width=.7\linewidth]{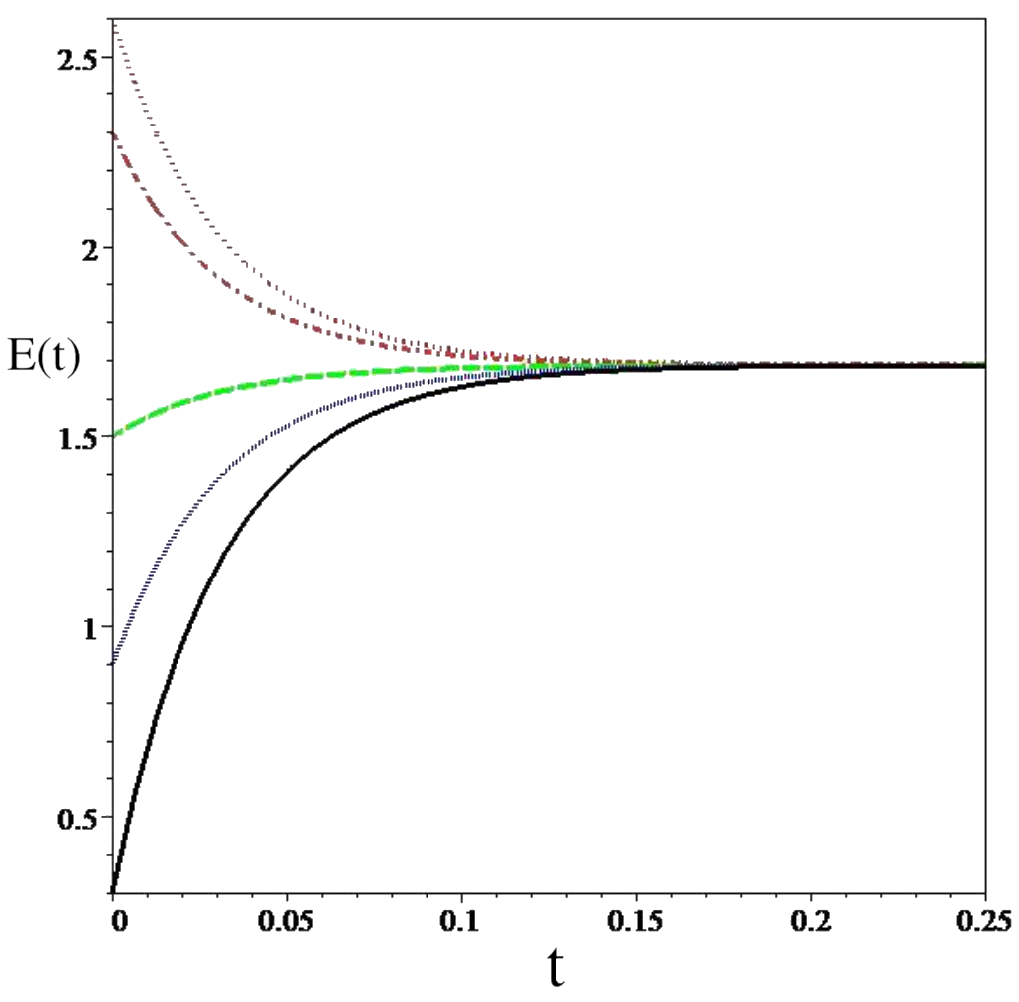}
\caption{Manipulation of the  quantum oscillator energy with algorithm (\ref{eq8}) for diferent initial conditions. Initial energy values are $E(0) = 0.3$ (black solid line); $E(0) = 0.9$ (blue dot line); $E(0) = 1.5$ (green dashed line); $E(0) = 2.3$ (orange dashdotted line); $E(0) = 2.6$ (red spacedotted line) to the target energy value $E_* = 1.8$. The other parameters are $\omega_0 = 1$,  $\gamma= 1$; $\Gamma_ 1 = 3$.
\label{fig4}}
\end{center}
\end{figure}

Dynamics of the incoherent only  energy control with the  algorithm (\ref{eq8}) for different initial conditions is shown in Fig.~\ref{fig4}.

\begin{figure}
\begin{center}
\includegraphics[width=1\linewidth]{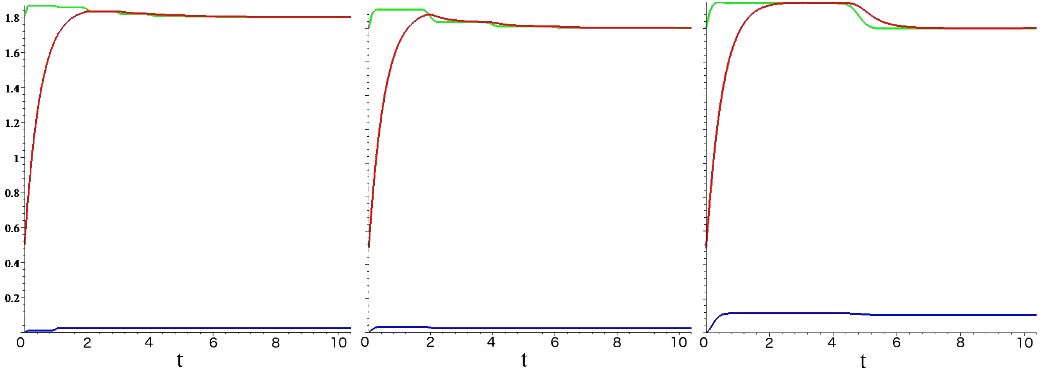}
\caption{Manipulation of the  quantum oscillator energy with sampled-data control algorithm. Left to right: sampling interval $h=1$, $h=2$, $h=5$. Red lines are for $E(t)$, green for $n(t)$, blue for $u(t)$.
The other parameters are the same as in Fig.~\ref{fig2}.
\label{fig5}}
\end{center}
\end{figure}

Figure~\ref{fig5} illustrates dynamics of sampled-data control systems for values of the sampling interval $h=1,2,5$. Robustification  parameter $\kappa$ was chosen as $\kappa=1.$ The rest parameters and initial conditions are the same. It is seen that the decrease of the system performance due to sampling is not too serious.

\section{Conclusions}
The problem of manipulation of energy in a quantum harmonic oscillator driven by coherent and incoherent controls is studied. Coherent field and mean number of excitations in the bath are used as control variables. General, differential (SGA-D) and finite (SGA-F)  forms of the speed gradient algorithm are developed. It is shown that SGA-D is applicable if both coherent and incoherent control are used while SGA-F is applicable when just incoherent control is used.

The differential form of the speed gradient algorithm is proved to be able to steer the energy of the oscillator to any prespecified target value. Conditions for non-negativity of spectral density obtained by the proposed method are found. As a result, not only the feasibility of the energy control in a quantum  harmonic oscillator is proven mathematically but also  a method to explicitly design physical coherent and incoherent controls which steer the energy of the oscillator to a prespecified value is provided.

An important problem of sampled-data control is also addressed. A robustified speed-gradient control algorithm in differential form is proposed. It is shown that the proposed robustified control algorithm ensures  exponential stability of the closed loop system which is preserved for sampled-data control.

Simulation results confirm reasonable performance of the closed loop systems. Degradation of the system performance due to sampling is not significant.

Future research may be aimed at extension of the proposed approach to more general and more complex classes of the controlled quantum systems.

\section*{Asknowledgements} Results of Section 3 (control algorithm design and analysis) were obtained in IPME RAS and supported by the Ministry of Science and Higher Education of the Russian Federation, project No. 075-15-2021-573. A.P. is partially supported in MISiS by the Ministry of Education and Science of the Russian Federation in the framework of the Program of Strategic Academic Leadership ”Priority 2030”.

\end{document}